# OPTIMIZATION OF A MU2E PRODUCTION SOLENOID HEAT AND RADIATION SHIELD USING MARS15*†

V.S. Pronskikh#, N.V. Mokhov, Fermilab, Batavia, IL 60510, U.S.A.

## Abstract

A Monte-Carlo study of several Mu2e Production Solenoid (PS) absorber (heat shield) versions using the MARS15 code has been performed. Optimizations for material as well as cost (amount of tungsten) have been carried out. Studied are such quantities as the number of displacements per atom (DPA) in the helium-cooled solenoid superconducting coils, power density and dynamic heat load in various parts of the PS and its surrounding structures. Prompt dose, residual dose, secondary particle flux are also simulated in the PS structures and the experimental hall. A preliminary choice of the PS absorber design is made on the ground of these studies.

*Work supported by Fermi Research Alliance, LLC under contract No. DE-AC02-07CH11359 with the U.S. Department of Energy.
†Presented paper at XX International Baldin Seminar on High Energy Physics Problems “Relativistic Nuclear Physics & Quantum Chromodynamics”, Dubna, October 4-9, 2010.
#vspron@fnal.gov

**OPTIMIZATION OF A MU2E PRODUCTION SOLENOID HEAT AND RADIATION SHIELD USING MARS15**

V.S. Pronskikh[1,2], N.V. Mokhov[2]
[1]Joint Institute for Nuclear Research, Dubna, 141980 Russia
[2]Fermi National Accelerator Laboratory, P.O.Box 500, Batavia 60510 IL, USA

Abstract
A Monte-Carlo study of several Mu2e [1] Production Solenoid (PS) absorber (heat shield) versions using the MARS15 [2] code has been performed. Optimizations for material as well as cost (amount of tungsten) have been carried out. Studied are such quantities as the number of displacements per atom (DPA) in the helium-cooled solenoid superconducting coils, power density and dynamic heat load in various parts of the PS and its surrounding structures. Prompt dose, residual dose, secondary particle flux are also simulated in the PS structures and the experimental hall. A preliminary choice of the PS absorber design is made on the ground of these studies.

## Introduction

The Mu2e experiment will be devoted to studies of the charged lepton flavor violation (CLFV) which up to now has never been observed and can manifest itself as the conversion of $\mu^-$ to $e^-$ in the field of a nucleus without emission of neutrinos. The emission of monoenergetic 105 MeV electrons can serve as a signature of such a process. CLFV has a very low probability ($<10^{-54}$) in the Standard Model, and its observation by the Mu2e [1] experiment can find its explanation in supersymmetric theories, extra dimensions, leptoquarks, compositeness, second Higgs doublet etc. Mu2e experiment is a further development of the proposed MECO [3] experiment.

One of the main parts of the Mu2e experimental setup is its production solenoid, in which negative pions are generated in interactions of the primary proton beam with the target (see Fig. 1). These pions then decay into muons which are delivered by the transport solenoid to the detectors. The off-axis 8 GeV proton beam will deliver $2 \cdot 10^{13}$ protons per second ($3 \cdot 10^{7}$ protons per pulse, every 1.7 μs) to the gold target, placed at the center of the PS bore.

## PS Absorber Material Optimization

The constraints in the PS absorber design are quench stability of the superconducting coils, low dynamic heat loads to the cryogenic system, reasonable lifetime of the coil components, acceptable hands-on maintenance conditions, compactness of the absorber that should fit into the PS bore and provide and aperture large enough to not compromise pion collection efficiency, cost, weight and other engineering constraints. The following ten versions have been studied in the course of PS absorber material optimization:

1) entirely tungsten absorber (Fig. 1), denoted #1 in Tables
2) five multilayer absorbers (Figure 2), composed of
    a. 5cm W, 20 cm Fe, 12 cm $BCH_2$, 3 cm Fe (from inside outside), #2
    b. 5cm W, 20 cm Fe, 12 cm $BCH_2$, 3 cm Cd, #3
    c. 5cm W, 20 cm Fe, 12 cm $BCH_2$, 3 cm Cu, #4
    d. 5cm W, 20 cm Cu, 12 cm $BCH_2$, 3 cm Fe, #5
    e. 5cm W, 20 cm WC, 12 cm $BCH_2$, 3 cm Fe, #6
3) entirely tungsten carbide (WC) absorber (Figure 1), #7
4) entirely $^{238}U$ absorber (Figure 1), #8
5) tungsten absorber with a conic copper part (Figure 3), #9
6) tungsten absorber with a cylindrical copper part (Figure 4), #10.

Tungsten carbide (WC) means a mix of 80% WC and 20% water. The composition of the multilayer absorbers was chosen according to their purposes: 1-st layer to stop charged particles, 2-nd layer to slow down the neutrons resulting from interactions in the target and 1-st layer, 3-rd layer to capture the slow neutrons and 4-th layer to suppress γ-quanta resulting from neutron capture.

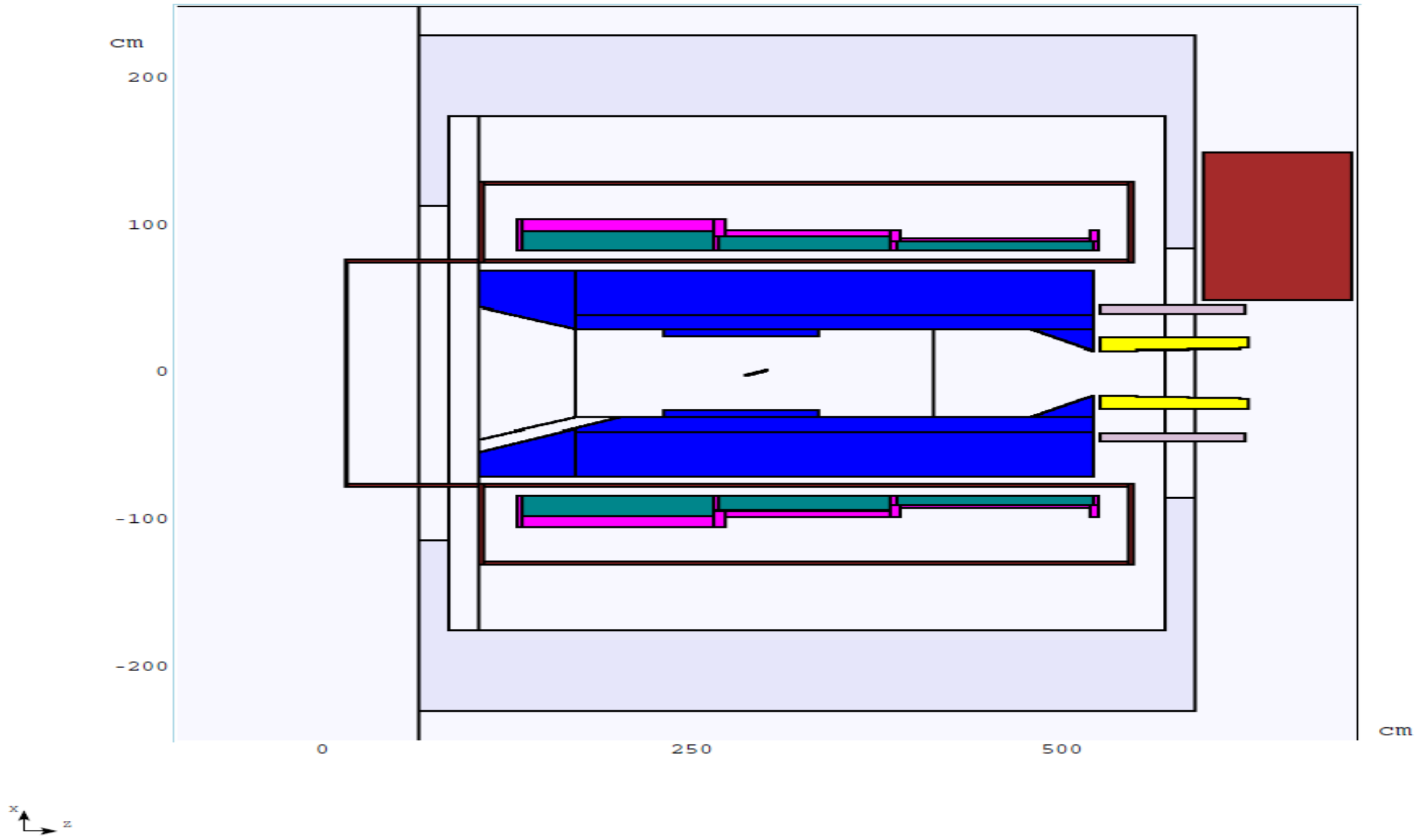

**Figure 1. Cross section of a tungsten (or any mono-material) absorber.**

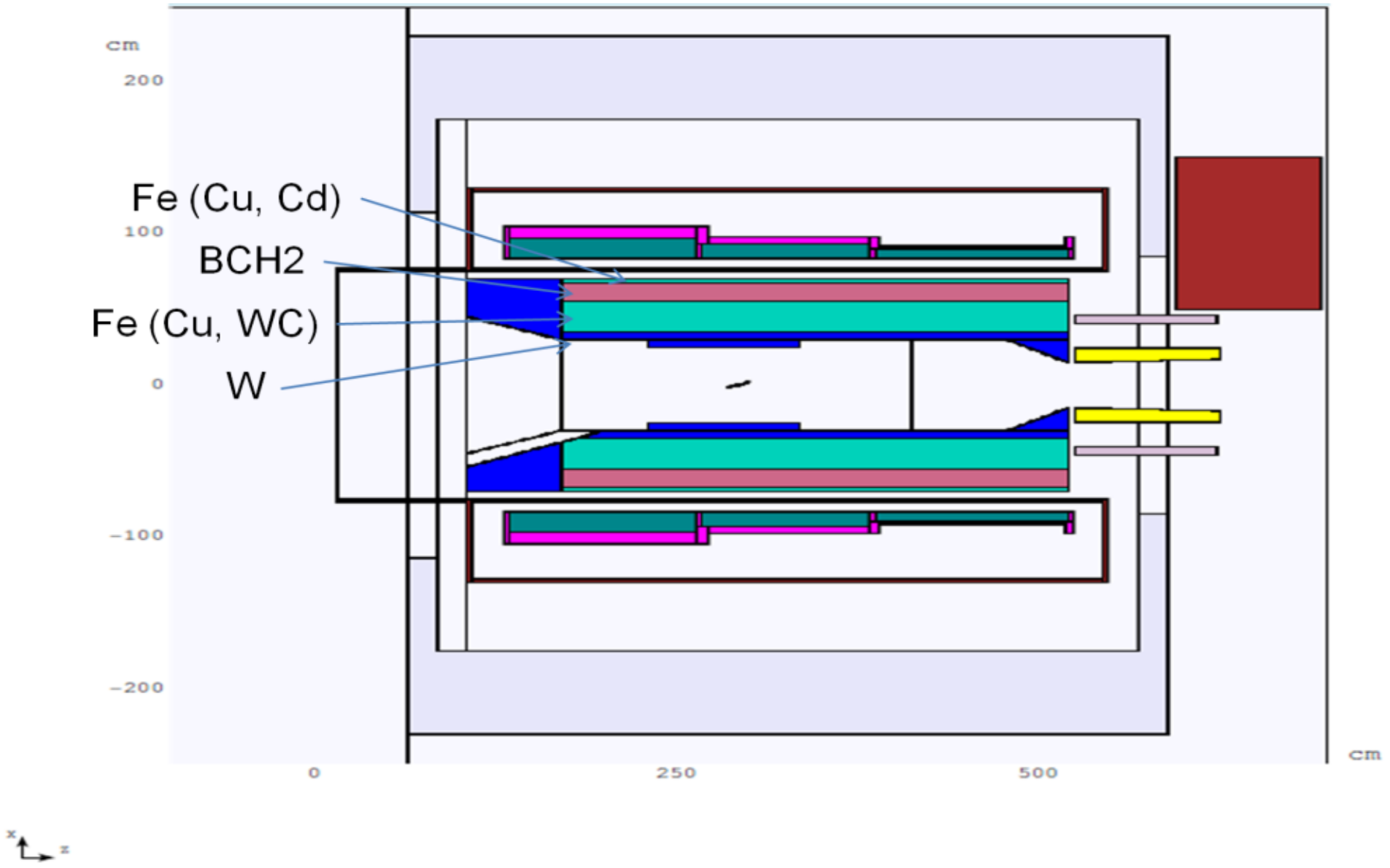

**Figure 2. Cross section of a multilayer absorber.**

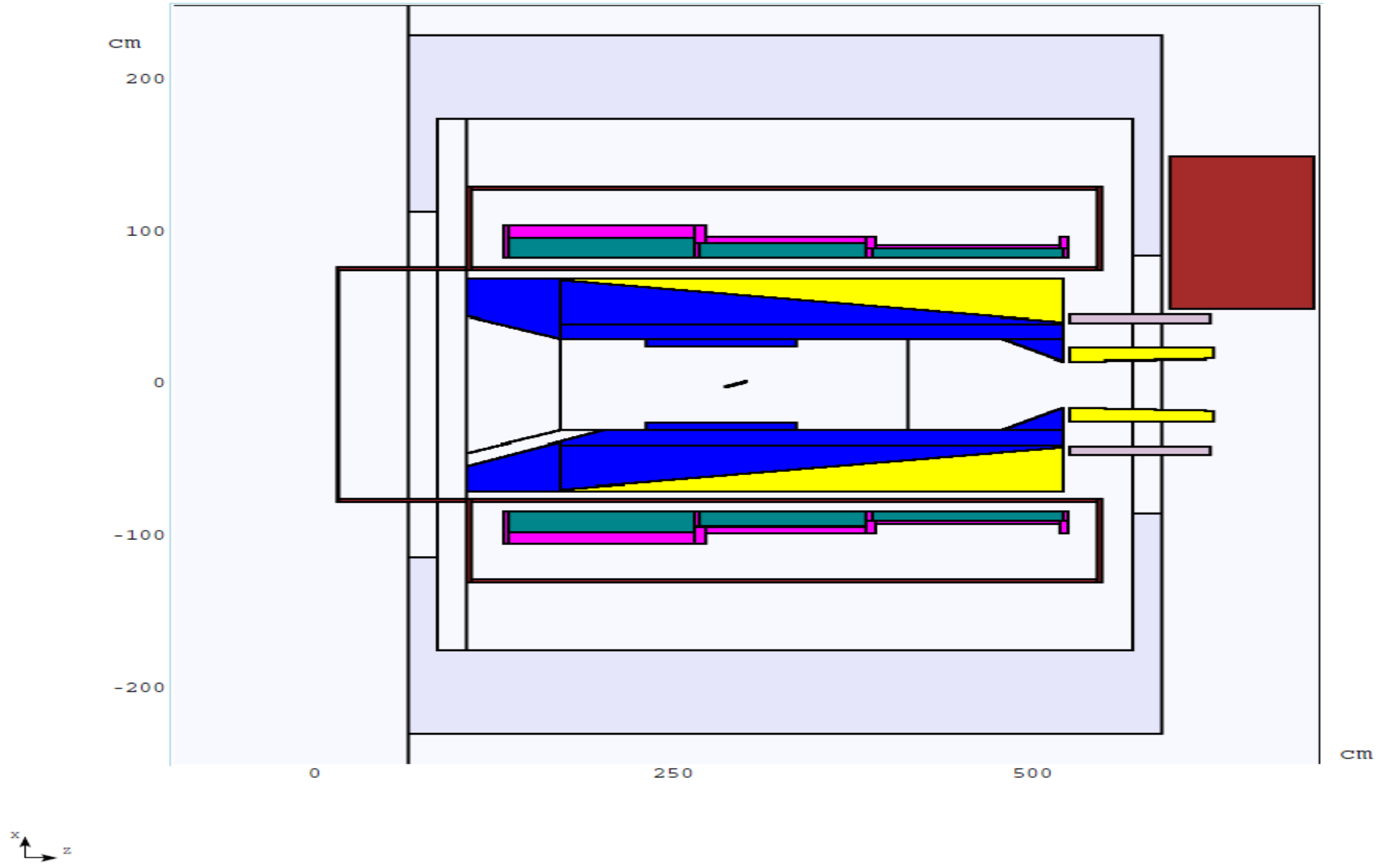

**Figure 3. Cross section of tungsten absorber with a conic copper part.**

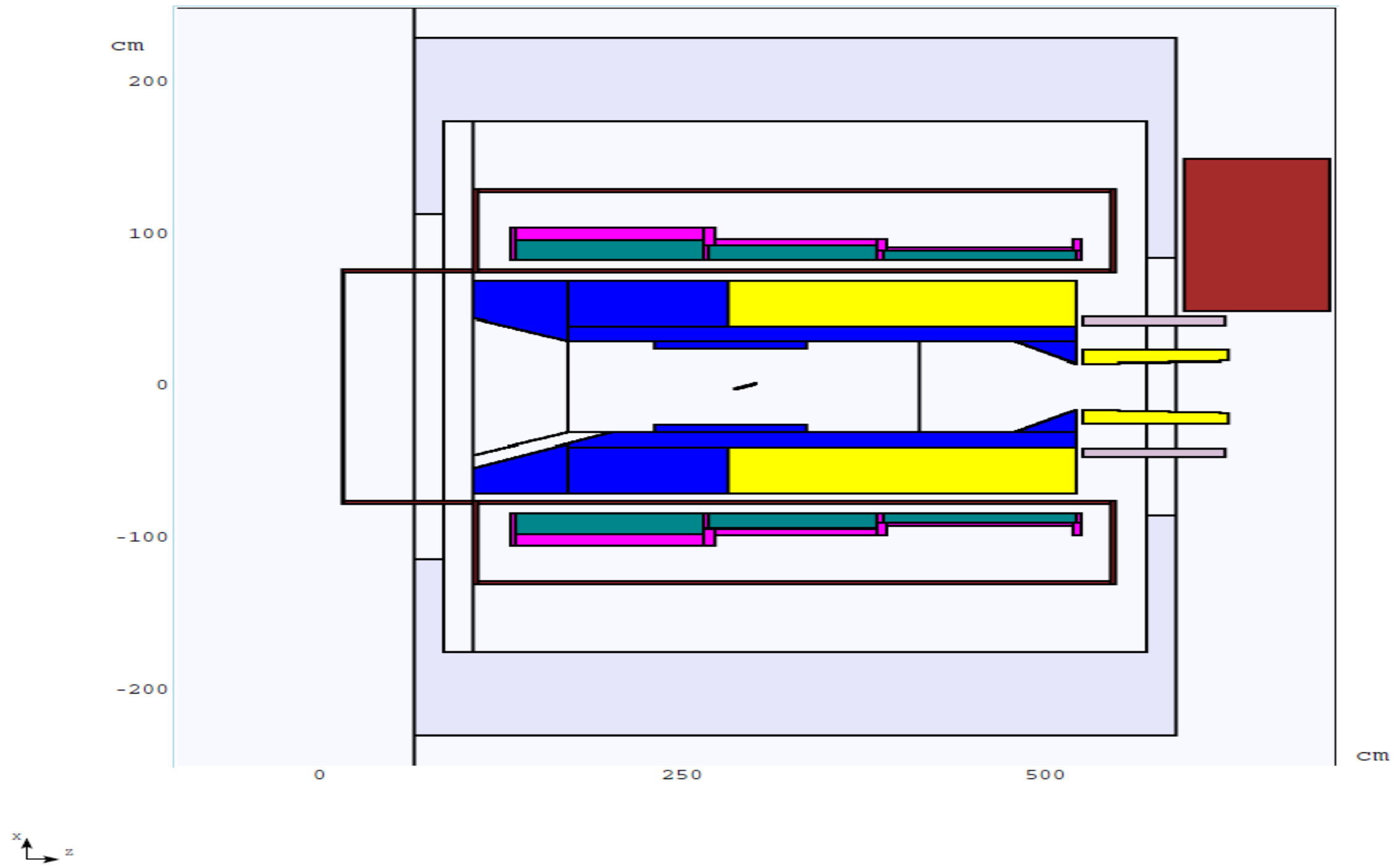

**Figure 4. Cross section of tungsten absorber with a cylindrical copper part.**

When optimizing the absorber, the following parameters were taken into account: dynamic heat load, peak power density, number of displacements per atom (DPA) in the helium-cooled solenoid coils, peak prompt dose and peak neutron flux in the superconducting coils. As one of the primary functions of the absorber is to protect the coils from warming and consequential quench, first two parameters serve to determine if the critical heating is attained and also to determine requirements to the cooling system.

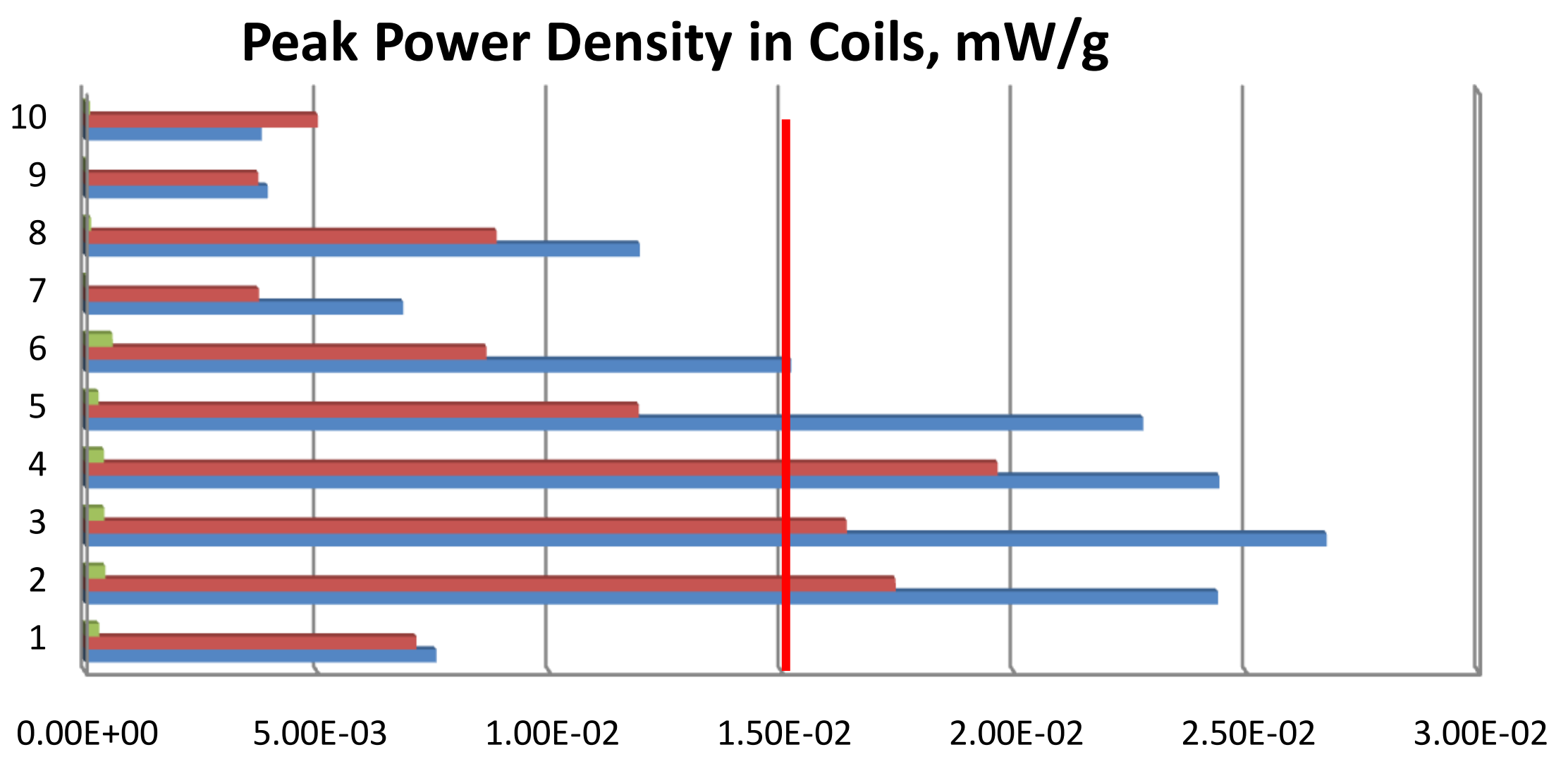

| | 1 | 2 | 3 | 4 | 5 | 6 | 7 | 8 | 9 | 10 |
|---|---|---|---|---|---|---|---|---|---|---|
| 3a | 3.04E-04 | 4.30E-04 | 4.10E-04 | 4.05E-04 | 2.89E-04 | 5.90E-04 | 4.86E-05 | 1.28E-04 | 7.00E-05 | 1.03E-04 |
| 2a | 7.13E-03 | 1.75E-02 | 1.64E-02 | 1.97E-02 | 1.19E-02 | 8.64E-03 | 3.74E-03 | 8.86E-03 | 3.73E-03 | 5.01E-03 |
| 1a | 7.56E-03 | 2.44E-02 | 2.67E-02 | 2.44E-02 | 2.28E-02 | 1.52E-02 | 6.85E-03 | 1.19E-02 | 3.93E-03 | 3.80E-03 |

**Figure 5. Peak power density in coils for the ten absorber versions, mW/g. (numbering see page 2).**

As one can see, the highest power densities are attained in the most absorbers in the first coil (only in #10 in the second), then in the second (#10 in the first), and the lowest (generally one or two orders of magnitude less) in the third. In all the considered absorbers except for the multilayer ones the power density values are not critical from the requirement point of view, however, the most promising from this point of view look absorbers #9 and #10.

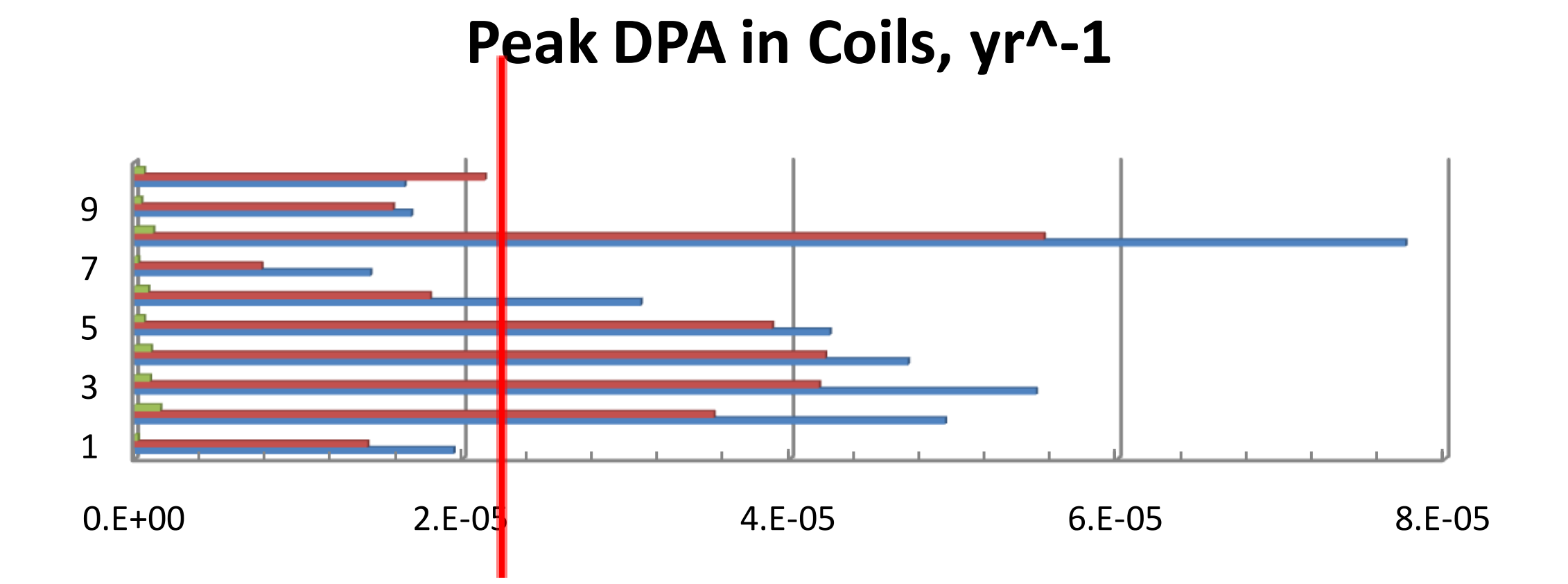

| | 1 | 2 | 3 | 4 | 5 | 6 | 7 | 8 | 9 | 10 |
|---|---|---|---|---|---|---|---|---|---|---|
| 3a | 2.23E-07 | 1.59E-06 | 9.62E-07 | 1.02E-06 | 5.94E-07 | 8.54E-07 | 2.42E-07 | 1.16E-06 | 4.18E-07 | 5.93E-07 |
| 2a | 1.42E-05 | 3.54E-05 | 4.18E-05 | 4.22E-05 | 3.89E-05 | 1.81E-05 | 7.77E-06 | 5.55E-05 | 1.58E-05 | 2.14E-05 |
| 1a | 1.95E-05 | 4.95E-05 | 5.50E-05 | 4.72E-05 | 4.25E-05 | 3.09E-05 | 1.44E-05 | 7.76E-05 | 1.69E-05 | 1.65E-05 |

**Figure 6. Peak DPA in coils for the ten absorber versions, per year.**

Like in the case of power density, peak DPA (Figure 6) are the highest in the first coil for all the absorbers except for #10, for which its highest value was obtained in the second coil. The second coil follows the first from the standpoint of the peak DPA value, and the third one shows the smallest value of it, by one or two orders of magnitude. As one can see in Figure 6, six of the absorbers (#2-6 and #8) exhibit DPA values higher than $2.5 \times 10^{-5}$ per year, whereas in the other four (entire tungsten #1, tungsten carbide with water #7, and two tungsten/copper absorbers #9 and #10) this value is below that limit [4].

## Conclusions

On the basis of material optimization studies, absorber based on W/Cu combination can be preliminary chosen as the best options among the considered ones.